\documentclass{article}
\usepackage[T1]{fontenc}
\usepackage[utf8]{inputenc}
\usepackage{ismir} 
\usepackage{amsmath,cite,url}
\usepackage{graphicx}
\usepackage{color}
\usepackage{mathabx}
\usepackage{subcaption}
\usepackage{booktabs}
\usepackage{caption}
\usepackage{enumitem}

\title{TOMI: Transforming and Organizing Music Ideas for Multi-Track Compositions with Full-Song Structure}

\multauthor
  {Qi He$^1$ \hspace{1cm} Gus Xia$^1$ \hspace{1cm} Ziyu Wang$^{1,2}$}
  {
  $^1$ Music X Lab, MBZUAI\hspace{1cm}
  $^2$ New York University\\
  {\tt\small heqi201255@icloud.com, gus.xia@mbzuai.ac.ae, ziyu.wang@nyu.edu}
  }
\def\authorname{Q. He, G. Xia, and Z. Wang}

\begin{document}

\maketitle

\begin{abstract}
Hierarchical planning is a powerful approach to model long sequences structurally. Aside from considering hierarchies in the temporal structure of music, this paper explores an even more important aspect: \textit{concept hierarchy}, which involves generating music ideas, transforming them, and ultimately organizing them—\textit{across musical time and space}—into a complete composition. To this end, we introduce \textbf{TOMI} (\textbf{T}ransforming and \textbf{O}rganizing \textbf{M}usic \textbf{I}deas) as a novel approach in deep music generation and develop a TOMI-based model via instruction-tuned foundation LLM. Formally, we represent a multi-track composition process via a sparse, four-dimensional space characterized by \textbf{clips} (short audio or MIDI segments), \textbf{sections} (temporal positions), \textbf{tracks} (instrument layers), and \textbf{transformations} (elaboration methods). Our model is capable of generating multi-track electronic music with full-song structure, and we further integrate the TOMI-based model with the REAPER digital audio workstation, enabling interactive human-AI co-creation. Experimental results demonstrate that our approach produces higher-quality electronic music with stronger structural coherence compared to baselines.\footnote{Source code: \url{https://github.com/heqi201255/TOMI}. Demo page: \url{https://tomi-2025.github.io/}.}

\end{abstract}

\section{Introduction}
Automatic music generation has advanced from producing short clips to composing entire pieces, yet long-term structure remains a major challenge. Unlike short-term generation, which focuses on capturing local patterns \cite{scg,polydiff,riffusion,deepbach,musiclm,musicgen}, long-term generation requires handling structure across multiple levels, from sectional repetition and cadence to the overall theme and whole narrative flow. The most common approach is to scale up models and data \cite{jukebox,yue}, yet even with vast training corpora, achieving truly structured compositions remains difficult. An alternative is to model the \textit{temporal hierarchy} of music, where it evolves at multiple time scales, with different model components capturing context dependencies at each level \cite{wholesonggen,musicframeworks,musicvae,jukebox}.

\begin{figure}
    \centering
    \includegraphics[alt={TOMI structure},width=0.99\linewidth]{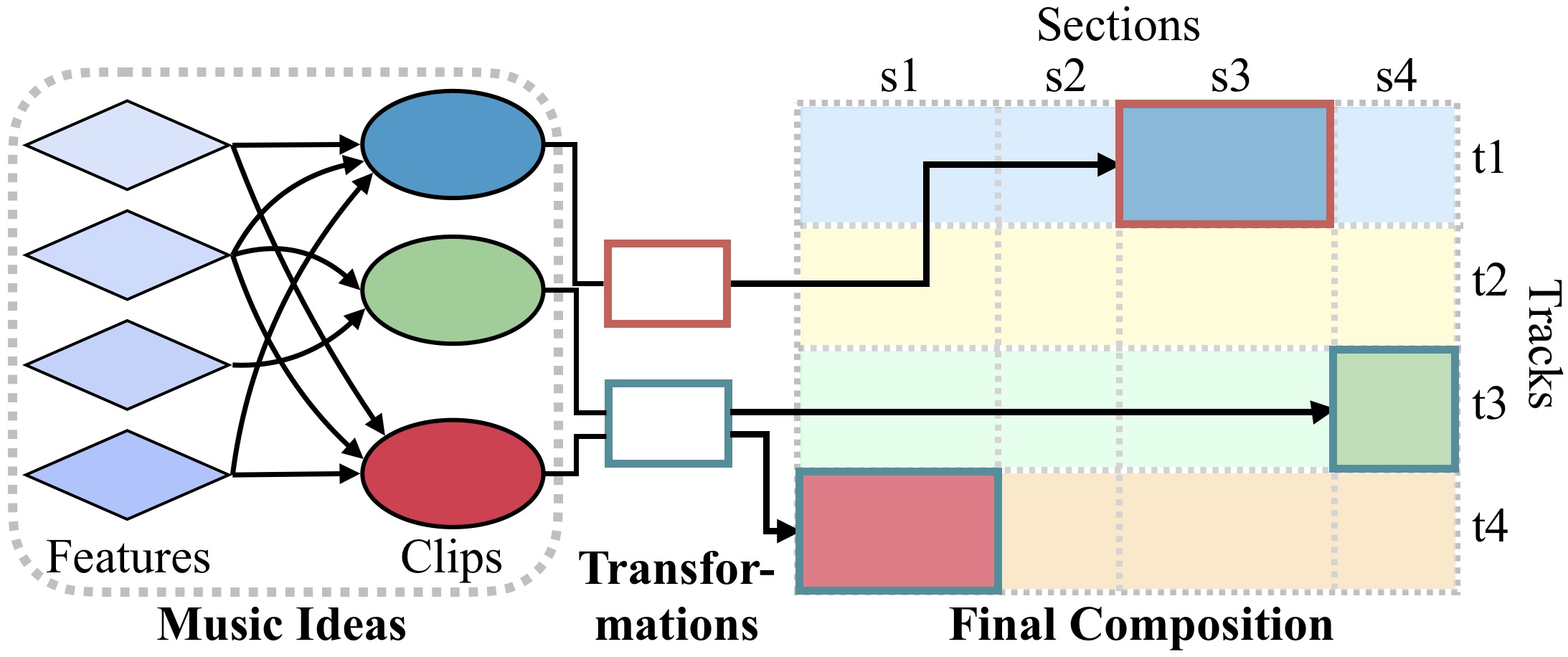}
    \caption{Concept hierarchy in TOMI: music ideas developed from features to clips are transformed and integrated into the composition, organized by sections and tracks.}
    \label{fig:fig1}
\end{figure}

But have we fully captured the hierarchical nature of music? While temporal hierarchy is essential, so is \textit{concept hierarchy}, which often manifests in the development of motifs and music materials. As noted in \cite{missingdeepmusic}, most pop songs are composed with a sparse and small set of core ideas, which are then evolved and repeated in an organized way. We illustrate this process in \figref{fig:fig1}, where music ideas, initially appearing as abstract features, are \textit{concretized} as music clips, then \textit{transformed}, and finally \textit{organized} into a full composition. We refer to this concept hierarchy in music as TOMI (\textbf{T}ransforming and \textbf{O}rganizing \textbf{M}usic \textbf{I}deas). TOMI shares a spirit with David Cope’s recombinant music composition in EMI \cite{emirecombinant}, an idea that deep learning has yet to truly embrace.

In this paper, we propose a TOMI-based music generation system for full-song and multi-track electronic music composition, inspired by the prevalent use of MIDI and audio sample packs among electronic music producers. The system is built around four key elements: \textit{clips}, \textit{transformations}, \textit{sections}, and \textit{tracks}. Sections and tracks are first created to serve as the canvas for composition, similar to a digital audio workstation (DAW) interface. The system then selects music clips from a library of audio and MIDI samples. For each selected clip, a transformation function is defined and applied, then the transformed clip is placed in its designated section and track. On the backend, a structured data representation parameterizes clips, transformations, sections, and tracks, and links them dynamically to construct a full composition. We leverage a pre-trained text-based large language model (LLM) to operate on this data structure, using in-context learning (ICL) to fill in the parameters and create dynamic links.

In sum, the contributions of this paper are as follows:

\setlength{\leftmargini}{\baselineskip}
\begin{enumerate}[topsep=0pt, noitemsep]
    \item \textbf{We introduce TOMI to model music concept hierarchy} and develop a deep learning-based system for structured electronic music generation. The proposed data structure integrates symbolic and audio representations and can be manipulated by text-based LLMs via ICL.
    \item \textbf{We apply our system to generate high-quality electronic music with full-song structure}. Objective and subjective evaluations show that songs generated by our model have clearer phrase boundaries, better phrase development, and higher music quality than the baselines.
    \item \textbf{We integrate TOMI with the REAPER digital audio workstation}, providing seamless connection with professional music software interface and enabling human-AI co-creation with high-resolution audio rendering.
\end{enumerate}

\section{Related Work}
In automatic music generation, many studies focus on generating coherent music segments \cite{scg,polydiff,riffusion,deepbach,musiclm,musicgen}, while fewer focus on modeling long-term structure under the temporal hierarchy of music. Jukebox \cite{jukebox} uses hierarchical VQ‑VAE with time conditioning to enhance long-term coherence; Wang \textit{et al}. \cite{wholesonggen} applies cascaded diffusion models for structured symbolic music generation. Some methods introduce explicit structure encoding in neural networks \cite{structureencodding, sympac} or use efficient sampling techniques to generate structured variations \cite{samplingvariations}. However, the application of TOMI concepts in music generation remains largely unexplored. Related works include a rule-based recombinant music method \cite{emirecombinant}, which reorganizes existing music elements based on stylistic constraints, and MELONS \cite{melons}, which uses a structure graph to enforce long-term dependencies in melody generation but offers limited flexibility in musical transformation.

In the context of co-creative music systems, existing generative models produce data in either symbolic (e.g., \cite{wholesonggen,polydiff,scg,melons,deepbach}) or waveform (e.g., \cite{riffusion,musiclm,musicgen}) format. Furthermore, these models lack intuitive interfaces for user co-creation and fine-grained control over musical elements, as found in modern DAWs. Composer's Assistant \cite{composersassistant} and its successor \cite{composersassistant2} integrate with REAPER to support co-creation but focus on generating symbolic phrases rather than full pieces. Our approach generates complete compositions containing both MIDI and audio phrases while also enabling user co-creation directly within REAPER.

With recent advances in AI, text-based LLMs have emerged as a promising alternative for music creation. Previous studies like ChatMusician \cite{chatmusician} and MuPT \cite{mupt} employ ABC Notation \cite{abcnotation} to represent music in text format and fine-tune LLaMA models \cite{llama}. While these approaches outperform generic LLMs in music tasks, their performance remains limited compared to LLMs trained exclusively on music data \cite{musecoco,musictransformer,antimusictransformer}. Other works have leveraged text-based LLMs for music analysis \cite{mullama,audioflamingo}, showing their capability to interpret music concepts through natural language. This motivates us to develop a textual hierarchical representation of music concepts to better integrate with text-based LLMs.

\begin{figure*}[ht]
    \centering
    \begin{subfigure}[t]{0.45\linewidth}
        \centering
        \includegraphics[width=\linewidth]{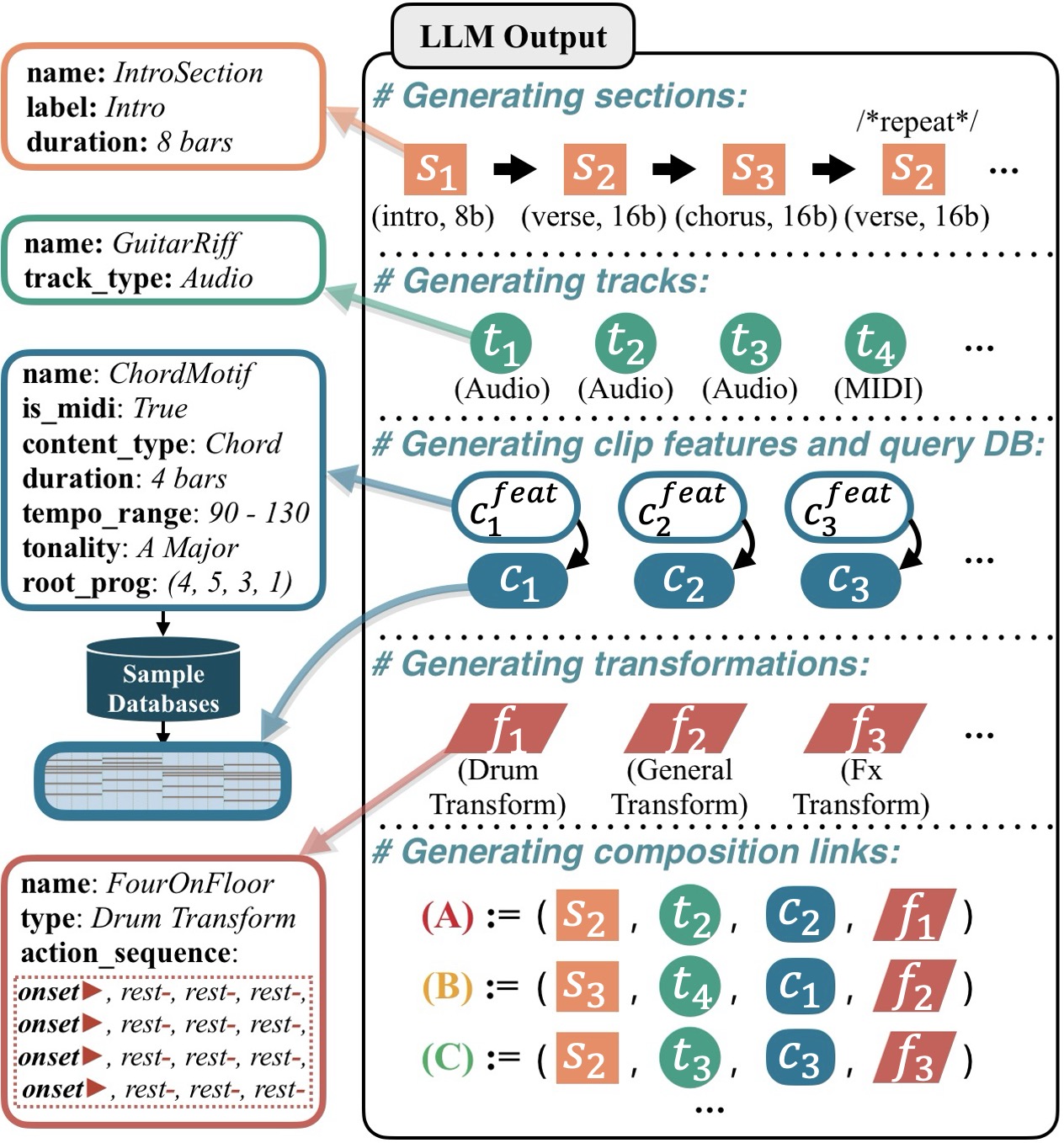}
        \caption{Example of an LLM output following our ICL prompt structure. In clips, the LLM only generates features for clips, which are then used to query the sample databases for actual MIDI or audio clips. We also show three composition links, each as a tuple of (section, track, clip, and transformation).}
        \label{fig:fig2a}
    \end{subfigure}
    \hfill
    \begin{subfigure}[t]{0.52\linewidth}
        \centering
        \includegraphics[width=\linewidth]{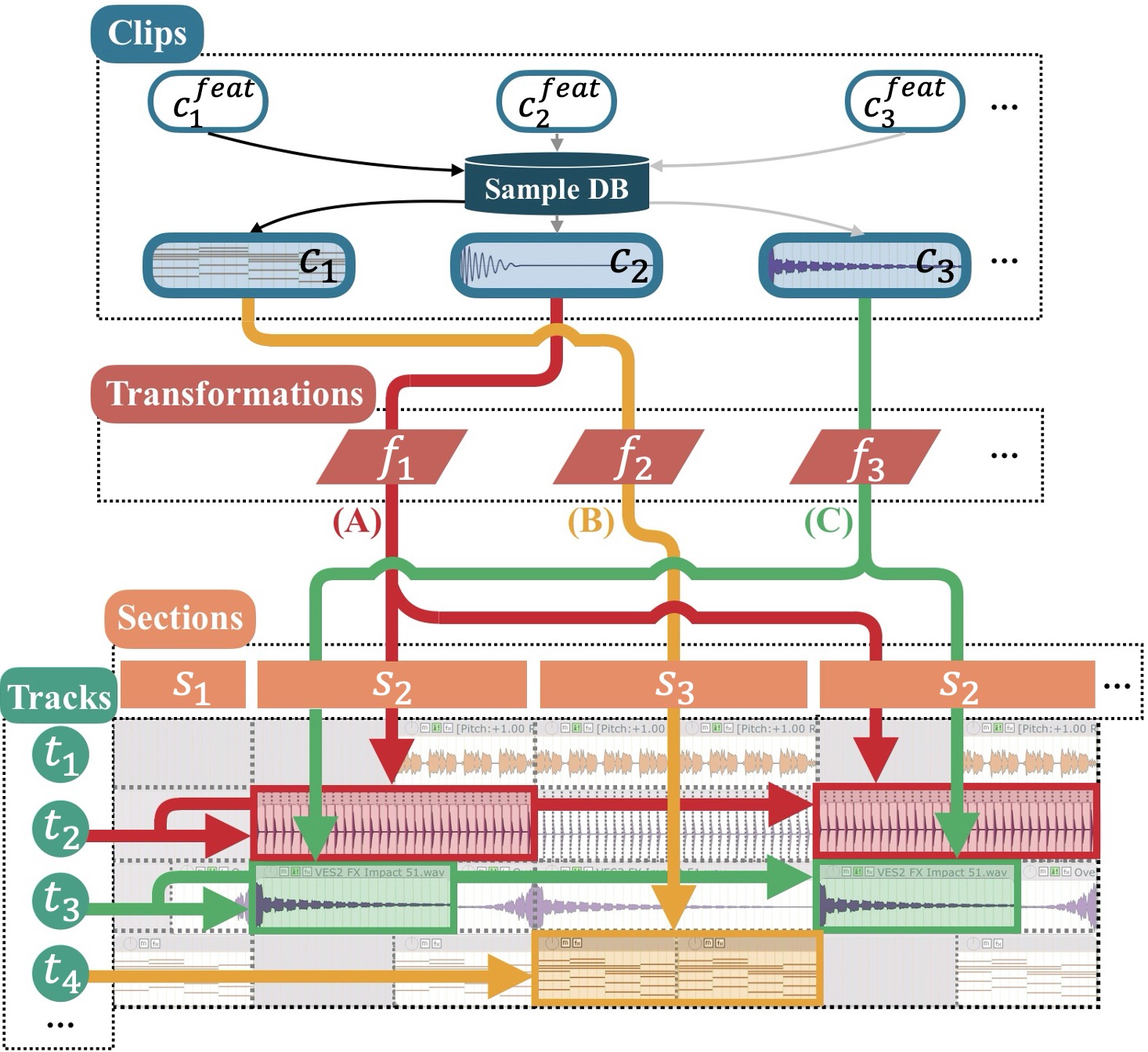}
        \caption{Music generation process with composition links, illustrating how clips (top) are transformed and then organized into the arrangement (bottom). Links $(A)$ and $(C)$ each have two branches because section $s_2$ appears twice in the section sequence, which means they are identical and share the same composition links.}
        \label{fig:fig2b}
    \end{subfigure}
    \caption{We show the structured LLM output in (a) with the corresponding music generation process in (b). We distinguish node types by color and shape, with a detailed attribute example for each type shown on the left of (a). We depict three composition link examples as colored arrows in (b), highlighting their results with rectangles in corresponding colors.}
    \label{fig:fig2}
\end{figure*}

\section{Methodology}
In this section, we discuss TOMI in multi-track electronic music generation with full-song structure. The implementation consists of two main components: (1) a graph data structure named \textit{composition link} that connects raw music ideas with the composition (Section~\ref{subsec:3:datastructure}), and (2) in-context learning to compose music by following this data structure (Section~\ref{subsec:3:icl}). We also demonstrate the integration with the REAPER DAW (Section~\ref{subsec:3:reaper}).

\subsection{TOMI Data Structure}
\label{subsec:3:datastructure}
The proposed data structure consists of four node types: \textit{clips}, \textit{sections}, \textit{tracks}, and \textit{transformations}. A \textit{composition link} is a quadruple of these nodes, specifying a music clip (\textit{what}) to be placed in a particular section (\textit{when}) and on a specific track (\textit{where}), undergoing certain transformations (\textit{how}). Nodes are reusable across links, forming a structured representation of the full composition.

\subsubsection{Clip Node}
Clips are audio or MIDI samples sourced from databases. Each clip is a short music segment, such as a chord progression or a drum loop, described by a set of features. A MIDI clip can represent elements like chords, basslines, melodies, or arpeggios, with attributes such as tonality, duration, and root progression. An audio clip can be either a sample loop or a one-shot sound, with keywords describing its content. The blue-edged box in \figref{fig:fig2a} shows an example feature set for a chord clip. Once the features are specified by the LLM (discussed in Section~\ref{subsec:3:icl}), we can query the databases for the corresponding music material.

\subsubsection{Section Node}
The temporal divisions of a music composition are modeled by section nodes. A section node involves a duration and a phrase label, such as verse and chorus. A section node can appear multiple times within a composition, meaning its music content remains identical across instances. For example, as shown in \figref{fig:fig2}, the 16-bar verse section $s_2$ is reused after the 16-bar chorus section $s_3$, resulting in two identical verses with the same content.

\subsubsection{Track Node}
Track nodes represent the vertical layering of a composition and organize clips into MIDI or audio tracks (see the track axis in \figref{fig:fig2b}). MIDI tracks accept only MIDI clips and require instruments to generate sound, while audio tracks accept only audio clips and play them directly.

\subsubsection{Transformation Node}
A transformation node transforms clips before placing them on specific tracks and sections. Unlike pitch transposition and tempo adjustment, which are handled in the final stage (discussed in Section~\ref{subsec:3:reaper}), transformation nodes perform more semantically meaningful manipulations and mainly serve three roles. First (and most importantly), it can control rhythmic patterns of the audio or MIDI clips with an \textit{action sequence} of onsets, sustains, and rests (see $f_1$ in \figref{fig:fig2a}). For example, it can convert MIDI chord clips with whole notes into rich rhythmic patterns or place a one-shot drum on each beat to create a four-on-the-floor rhythm. Second, it can handle the riser or faller sound effects commonly used in electronic music composition, which are placed either left- or right-aligned within a section, creating smooth transitions (see $c_3$ in \figref{fig:fig2b}). Lastly, it dynamically handles the looping of a clip, determining whether it plays once at a specific time, loops throughout a section, or is trimmed to fit a shorter section. Our implementation defines three subclasses of transformations: (1) Drum transform (for one-shot drums), (2) Fx transform (for risers and fallers), and (3) General transform (for others), each following different transformation rules. We refer the reader to our demo page for details.

\subsubsection{Composition Link}
A composition link comprises one node each of section, clip, transformation, and track, showing the entire process of transforming and organizing a music idea in the composition. Since nodes are independent of the composition link, they can be reused across multiple links. \figref{fig:fig2b} shows the process of three composition links. Note that links $(A)$ and $(C)$ branches in the final composition because section $s_2$ is used twice. This is an efficient way to represent complex arrangements, as a single clip can be referenced by multiple links, 
spanning different sections and tracks while adapting to various transformations.

\subsection{Music Generation with In-Context Learning}
\label{subsec:3:icl}
The TOMI data structure, as defined above, can be fully represented in text format. A complete composition can be decomposed into a set of composition links, each represented as a quadruple of nodes, where each node corresponds to a set of textual attributes. By leveraging a text-based LLM with in-context learning, we can generate compositions directly as TOMI instances. Our ICL prompt systematically breaks down the data structure in steps, following the order: \textit{sections}, \textit{tracks}, \textit{clips}, \textit{transformations}, and \textit{composition links}. It guides the LLM from 
planning the overall song structure and music ideas to organizing them into a full song. 
The LLM output follows these steps accordingly, as shown in \figref{fig:fig2a}.

In our implementation, we elaborate our data structure with detailed examples and define a structured response schema in the prompt. We require the LLM to generate a unique \textit{name} attribute for each node to reference in composition links. Moreover, we can specify additional contexts, such as tempo, mood, and custom song structures. To get robust results, we implement a rule-based validation to check the LLM output for syntax errors and invalid values. If issues are detected, an error report is generated, prompting the LLM to iteratively refine its output until no errors remain. At this point, we obtain an abstract representation of the composition in TOMI structure. To realize this, we initiate the sample retrieval process to get the actual clip materials. Then, we set a global \textit{tempo} and \textit{key} to unify the keys and tempos of clips within the DAW.

\subsection{Digital Audio Workstation Integration}
\label{subsec:3:reaper}
To support audio rendering and interactivity, we integrate the TOMI framework with the REAPER DAW. This allows the generated composition to be visualized in a professional DAW while benefiting users from REAPER’s editing and rendering capabilities. The composition links and nodes of a generated piece are converted to REAPER elements, including tracks, section markers, and clips with applied transformations. We use REAPER to automatically time-stretch loopable clips to fit the \textit{tempo} setting and transpose melodic clips to align with the \textit{key} setting.

\section{Experiment}
To implement the generation system, we prepare a MIDI database and an audio database for clip sample retrieval and use GPT-4o \cite{gpt4o} to generate compositions in TOMI schema. We evaluate our approach with baseline methods and use both objective and subjective measurements to compare the music quality and structural consistency.

\subsection{System Preparation}
\label{subsec:4:database}
We collect multiple licensed MIDI and audio sample packs in the electronic music genre through online purchases.\footnote{We obtain sample packs from two music asset platforms: \url{https://splice.com/} and \url{https://www.loopmasters.com/}} We process the raw datasets into separate MIDI and audio databases using different feature extraction methods. For MIDI clips, we developed a script to analyze and extract musical features from them. Moreover, it can also extract music stems, such as bass, chord, and melody, from the source MIDI to augment the data. Then, we store the labeled data in a SQLite3 \cite{sqlite} database as our MIDI database. For audio samples, we use ADSR Sample Manager \cite{adsr} to analyze and generate labels for them. The results are also exported as a SQLite3 database. The statistics of our MIDI database and audio database are shown in \tabref{tab:databases}. Sample retrieval involves constructing a search query based on the clip's attributes to fetch matching samples from the database. The clip node randomly selects one sample from the results. If no matches are found, the clip and its associated composition links are discarded.

We limit all generated sections to the 4/4 time signature to simplify implementation. When exporting audio via REAPER, we randomly assign each MIDI track to one of eight virtual instrument presets (5 for chords, 2 for melody, and 1 for bass). We keep all REAPER settings at their defaults and apply no mixing plug-ins except for a limiter on the master track to prevent audio clipping.

\begin{table}
    \centering
    \small
    \subfloat[MIDI content types.]{
        \begin{tabular}{lr}
            \hline
            Content Type & Count \\
            \hline
            Chord & 2604        \\
            Bass & 209        \\
            Melody & 1392        \\
            Arpeggio & 227        \\
            \textbf{Total} & \textbf{4432} \\
            \hline
        \end{tabular}
    } \quad 
    \subfloat[MIDI durations.]{
        \begin{tabular}{lr}
            \hline
            Duration & Count \\
            \hline
            4-bar & 2947        \\
            8-bar & 1417        \\
            16-bar & 68        \\
            & \\
            \textbf{Total} & \textbf{4432} \\
            \hline
        \end{tabular}
    }
    \vspace{\baselineskip}
    \subfloat[Audio sample types.]{
        \begin{tabular}{lr}
            \hline
            Sample Type & Count \\
            \hline
            Loop & 104493        \\
            One-Shot & 170187        \\
            & \\
            & \\
            \textbf{Total} & \textbf{274680} \\
            \hline
        \end{tabular}
    } \quad 
    \subfloat[Audio loop durations.]{
        \begin{tabular}{lr}
            \hline
            Loop Duration & Count \\
            \hline
            2-bar & 24922        \\
            4-bar & 27214        \\
            8-bar & 24638        \\
            16-bar & 27719        \\
            \textbf{Total} & \textbf{104493} \\
            \hline
        \end{tabular}
    }
    \caption{Statistics of sample databases.}
    \label{tab:databases}
\end{table}

\subsection{Baseline Method and Ablations}
\label{subsec:4:baseline}
We compare our approach with MusicGen \cite{musicgen} and two ablations in electronic music generation. Since our system integrates both audio and MIDI data, there is no directly comparable baseline. However, as our final output is rendered as audio, MusicGen is a suitable comparison, which is capable of generating long-term and high-quality electronic music. This allows us to evaluate our method against state-of-the-art music generation systems. The ablations help us assess the individual contributions of the composition link representation and the LLM integration.

\begin{description}[noitemsep]
    \item[MusicGen] We use the MusicGen-Large-3.3B model as the baseline with prompts specifying key, tempo, and section sequence. To generate longer audio beyond its duration limit, we apply a sliding window approach, where a 30-second window slides in 10-second chunks, using the previous 20 seconds as context. We modify its inference process to include the current generation's position within the full composition and its corresponding phrase notations in the prompt at each step, guiding the model to align its output with the given structure.
    \item[Standalone LLM (TOMI w/o Composition Links)] We remove the composition links representation from our system. We redesign the prompt to let the LLM generate a sequence of tracks and clip descriptions with position information (time point and track location) conditioned on a section sequence. The sample retrieval mechanism is also applied for clips.
    \item[Random (TOMI w/o LLM)] We replace the LLM operations in our system with a rule-based method that uses randomized operations to generate music within the composition-links structure. 
    The system creates 15–25 track nodes, populates sections with clips stochastically, and determines for each track whether to place, reuse, or generate a new clip. MIDI clips are assigned a random type (chord, bass, or melody) with bass derived from chords, while audio clips are selected from tonal, percussion, and sound effect feature labels. Each clip is then linked to one of four predefined transformations: general, drum, riser Fx, or faller Fx.
\end{description}
We define 4 keys and 4 distinct section sequences, each consisting of 8 to 10 sections. Each section has a name, a phrase label, and a duration ranging from 4 to 16 bars. 
Then, using each method, we generate 4 sets of electronic music pieces at 120 BPM, with each set containing 8 pieces (2 per key), all conditioned on the same section sequence. In total, we generate 32 compositions for each method.

\begin{table*}
    \centering
    \small
    \begin{tabular}{lrrrrr}
        \toprule
        Method & $\text{FAD}^{\mathrm{VGGish}} \downarrow$ & $\text{FAD}^{\mathrm{CLAP}} \downarrow$ & $\text{ILS}^{\mathrm{MERT}} \uparrow$ & $\text{ILS}^{\mathrm{MS}} \uparrow$ & $\text{ILS}^{\mathrm{WF}} \uparrow$ \\
        \midrule
        TOMI & \textbf{3.51} & \textbf{0.38} & \textbf{0.28} $\pm$ 0.12 & \textbf{0.36} $\pm$ 0.33 & \textbf{1.14} $\pm$ 0.73 \\
        MusicGen & 5.31 & 0.62 & 0.06 $\pm$ 0.04 & 0.12 $\pm$ 0.07 & 0.28 $\pm$ 0.09 \\
        Standalone LLM & 5.84 & 0.46 & 0.16 $\pm$ 0.11 & 0.10 $\pm$ 0.12 & 0.09 $\pm$ 0.16 \\
        Random & 6.92 & 0.47 & 0.16 $\pm$ 0.09 & 0.22 $\pm$ 0.16 & 0.48 $\pm$ 0.28 \\
        \bottomrule
    \end{tabular}
    \caption{Objective evaluation results of FAD with two models and ILS with three latent representations.}
    \label{tab:objective_eval}
\end{table*}

\begin{figure*}
    \centering
    \includegraphics[alt={ILS example image},width=0.9\linewidth]{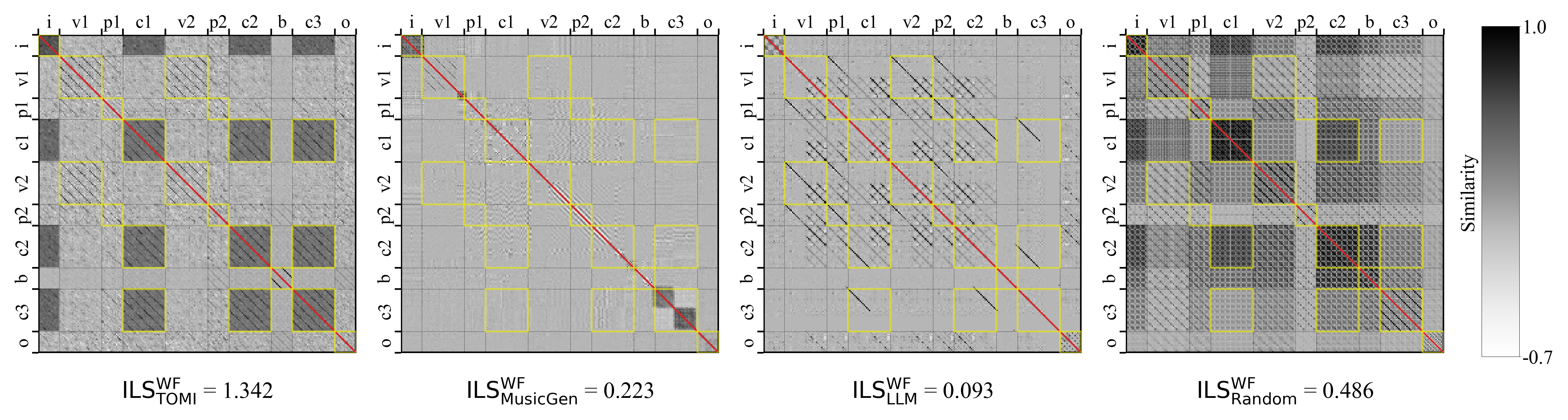}
    \caption{ILS similarity matrices example, where all four compositions are generated under the section sequence: \textit{i} (intro), \textit{v1} (verse 1), \textit{p1} (pre-chorus 1), \textit{c1} (chorus 1), \textit{v2} (verse 2), \textit{p2} (pre-chorus 2), \textit{c2} (chorus 2), \textit{b} (bridge), \textit{c3} (chorus 3), and \textit{o} (outro). Darker colors indicate higher segment similarity. The blocks marked as yellow-edge boxes are same-label similarities, with diagonal values marked as red lines being excluded, and the remaining parts are different-label similarities.}
    \label{fig:ils}
\end{figure*}

\subsection{Objective Evaluation}
We evaluate the music quality and structural consistency of generated compositions. For music quality, we use the \textbf{Fréchet Audio Distance} metric (FAD) \cite{fad,fad_pytorch} with a VGGish model \cite{vggish} and a CLAP model \cite{clap} to compare human-composed electronic music and the music generated by each method. We randomly collected 329 songs as references from the Spotify Mint playlist \cite{spotify}, one of the most popular curated playlists dedicated to the electronic music genre. A lower FAD score indicates that the generated music is closer to human-composed music in quality.

For structural consistency, we use the \textbf{Inter-Phrase Latent Similarity} metric (ILS) refined from \cite{wholesonggen}. ILS aims to compute a self-similarity matrix of musical features and evaluates if the average similarity between segments sharing the same phrase label is higher than those with different labels. Instead of measuring the ratio of same-label to overall similarities in the original metric, we use Cohen's \textit{d} \cite{cohensd} to compute the effect size of the difference between same-label and different-label similarities, excluding diagonal elements to avoid biases. This offers a scale-independent measure that robustly captures the separation between groups. We extract latent representations from audio while preserving temporal structure, then compute a self-similarity matrix of them using cosine similarity. We evaluate ILS with MERT embeddings \cite{mert}, Mel spectrograms, and raw waveforms, denoted as $\text{ILS}^{\mathrm{MERT}}$, $\text{ILS}^{\mathrm{MS}}$, and $\text{ILS}^{\mathrm{WF}}$, respectively. To compute the ILS score, we denote the cosine similarity between time-related latent elements $i$ and $j$ as $S_{ij}$, the phrase label of element $i$ as $l_i$, and the quantities of same-label and different-label elements as $N_\mathrm{same}$ and $N_\mathrm{diff}$, respectively, excluding diagonal elements. A higher ILS score indicates better and more effective structural consistency. The formula is defined as:

\begin{equation}
\text{ILS} = \frac{\bar{X}_\mathrm{same} - \bar{X}_\mathrm{diff}}{s}\text{,}
\end{equation}
where $\bar{X}_\mathrm{same}$ and $\bar{X}_\mathrm{diff}$ are the mean similarities of same-label and different-label groups, respectively, defined as:
\begin{align}
\bar{X}_\mathrm{same} &= \frac{\sum_{i\neq j} S_{ij} \cdot \delta(l_i = l_j)}{N_\mathrm{same}} \text{,} \\
\bar{X}_\mathrm{diff} &= \frac{\sum_{i, j} S_{ij} \cdot \delta(l_i \neq l_j)}{N_\mathrm{diff}}\text{.}
\end{align}
and $s$ is the pooled standard deviation derived from the variances $s_\mathrm{same}^2$ and $s_\mathrm{diff}^2$ of the two groups, defined as:

\begin{equation}
s = \sqrt{\frac{(N_\mathrm{same} - 1)s_\mathrm{same}^2 + (N_\mathrm{diff} - 1)s_\mathrm{diff}^2}{N_\mathrm{same} + N_\mathrm{diff} - 2}}
\end{equation}

We show the FAD scores and ILS with means and standard deviations in \tabref{tab:objective_eval}. 
The results show that our method achieves the lowest FAD scores and the highest ILS scores across all measurements, outperforming the baseline methods in both music quality and structural consistency. Notably, the ablation results show that even with high-quality music samples, lacking sufficient arrangement logic can still lead to poor FAD scores. This proves our approach's ability to generate music with improved flow naturalness and arrangement coherence while preserving the expected long-term structures. 
\figref{fig:ils} provides a comparative visualization of ILS matrices for four compositions, with phrase labels on both axes, yellow-edged boxes highlighting regions of same-label similarities, the red line marking the main diagonal, and the remaining areas representing different-label similarities.

\begin{figure*}
    \centering
    \includegraphics[alt={subjective eval result image},width=0.7\linewidth]{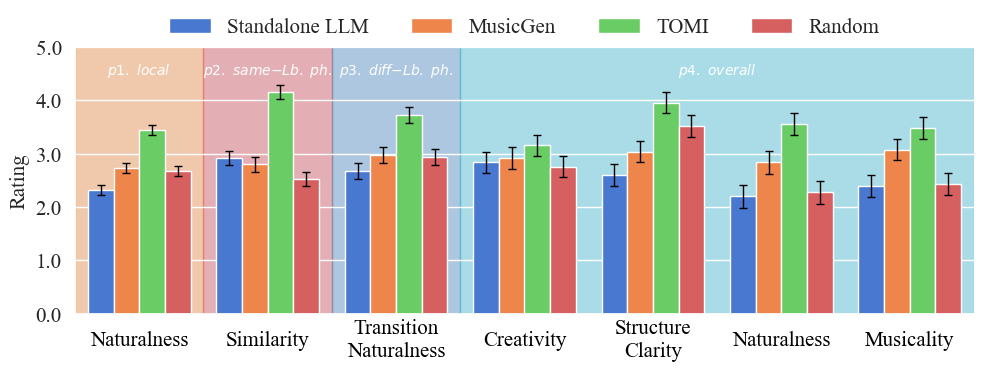}
    \caption{Subjective evaluation results of mean score with within-subject confidence intervals, where \textit{p1}-\textit{p4} corresponds to four parts of our survey: \textit{p1. Local Music Quality}, \textit{p2. Consistency Among Same-Label Phrases}, \textit{p3. Contrast Between Different-Label Phrases}, and \textit{p4. Overall Full-Song Evaluation}.}
    \label{fig:subjective_eval}
\end{figure*}

\subsection{Subjective Evaluation}
We conduct a double-blind online survey to compare the music quality and structural consistency of compositions generated by the four methods. Since each composition is a full 3- to 4-minute song, which is too long to be evaluated in a single-question, we assess four compositions (one per method) throughout the survey. To ensure a comprehensive evaluation, we create three distinct question sets with the same survey structure but different compositions, resulting in a total of 12 compositions being evaluated. 
The survey is divided into four parts, each containing 4 to 12 questions, aiming to evaluate the piece from section-level music quality to overall structural alignment. All metrics are measured on a 5-point rating scale. The details are as follows:
\setlength{\topsep}{0pt}
\begin{description}[noitemsep]
    \item[Part 1. Local Music Quality] This part consists of 3 subparts, each selecting the same section (e.g., intro) from each composition. Participants rate \textit{Naturalness} to evaluate the similarity to human-composed music and the conformity to the typical electronic music style. 
    \item[Part 2. Consistency Among Same-Label Phrases] This part consists of 2 subparts, each selecting two sections with the same phrase label (e.g., verse 1 and verse 2) from each composition. Participants rate \textit{Similarity} between the two sections. 
    \item[Part 3. Contrast Between Different-Label Phrases] This part consists of 2 subparts, each selecting two consecutive sections from the same position in each composition (e.g., intro and verse 1). Participants rate \textit{Transition Naturalness} based on boundary clearness, transition smoothness, and phrase alignment.
    \item[Part 4. Overall Full-Song Evaluation] This part shows the complete composition audios. Participants rate \textit{Structure Clarity} for how well each section aligns with the given structure, \textit{Creativity} for how creative the music is, \textit{Naturalness} for how humanlike the music sounds, and \textit{Musicality} for the overall music quality. 
\end{description}

We insert an intermediate page between evaluation parts to inform participants of their progress and help reduce listening fatigue. Additionally, we show the input conditions for music generation on each question page to remind participants of the expected song structure and tonality. We distributed the survey on multiple social media platforms and received a total of 73 responses. 
The demographic statistics of participants are as follows:

\textit{Age} ($<$18: 0\%, 18-29: 69.86\%, 30-44: 19.18\%, 45-59: 5.48\%, $\geq$60: 5.48\%);

\textit{Gender} (Female: 30.14\%, Male: 65.75\%, Non-binary: 2.74\%, Prefer not to say: 1.37\%);

\textit{Music background} (Amateur: 34.25\%, Intermediate: 41.1\%, Professional: 24.66\%);

\textit{Years spent on studying music} (None: 26.03\%, 1 year: 6.85\%, 2 years: 10.96\%, 3-5 years: 13.7\%, 6-10 years: 16.44\%, $>$10 years: 26.03\%).

The subjective evaluation results are shown in \figref{fig:subjective_eval}, where the bar height represents the mean score, and the error bar represents the confidence intervals computed by within-subject ANOVA. The results show that our method significantly outperforms the baseline in most subjective metrics, proving the effectiveness of our system in generating high-quality electronic music with solid long-term structural consistency.

\section{Conclusion and Future Work}

We contribute TOMI, a concept hierarchy paradigm for music representation, and combine it with an ICL approach to achieve the first system for generating long-term, multi-track electronic music with both MIDI and audio clips. Experimental results show that our approach achieves high-quality generation with robust structural consistency. In addition, we integrate it with REAPER to support audio rendering and co-creation. However, harmonic coherence in our results can occasionally be disrupted by randomness and limited features during sample retrieval. Our system currently selects samples from local collections using a small feature set, which can lead to empty results or highly divergent samples. To improve this, we plan to integrate generative models for clips or use ML-based embedding models for MIDI and audio. Then, we aim to extend our model with a more sophisticated structural hierarchy to support advanced sound design and mixing. Lastly, we plan to train a TOMI-based neural network on music project files to enhance generation quality and scalability.

\bibliography{TOMI_refs}
\end{document}